\magnification=1200 	





\let\Malpha=\alpha
\def\alpha{\relax\ifmmode{\Malpha}\else{$\Malpha$}\fi}
\let\Mbeta=\beta
\def\beta{\relax\ifmmode{\Mbeta}\else{$\Mbeta$}\fi}
\let\Mgamma=\gamma
\def\gamma{\relax\ifmmode{\Mgamma}\else{$\Mgamma$}\fi}
\let\Msection=\section
\def\section{\relax\ifmmode{\Msection}\else{$\Msection$}\fi}
\let\Mdollar=\$
\def\${\relax\ifmmode{\Mdollar}\else{$\Mdollar$}\fi}
\let\Ldots=\ldots
\def\ldots{\relax\ifmmode{\Ldots}\else{$\Ldots$}\fi}


\def\newline{\hfil\break}
\def\indent{\hbox to 20pt{}}

\def\doublespace{\baselineskip 24 pt}

\def\blankline{\par\vskip 12 pt\noindent}


\newcount\listno
\def\beginlist{\listno=0}
  \beginlist
\def\list{\par\noindent\hangindent 20pt\advance\listno by 1 
  \hbox to 20pt{\hfil\relax\the\listno.\ }}

\newcount\listsno
\def\beginlists{\listsno=0}
  \beginlists
\def\lists{\par\noindent\hangindent 20pt\advance\listsno by 1 
  $^{\relax\the\listsno}$}

\def\ref{\par\noindent\hangindent 20pt}

\def\nobreak{\penalty1000}
\def\titl#1\endtitl{\par\vfil\eject
     \par\vbox to 2in {}{\bf #1}\par\vskip 1in\nobreak}
\def\sect#1\endsect{\par\vfil\eject{\bf #1}\par\vskip 12pt\nobreak}
\def\subsect#1\endsubsect{\par
      \vskip 0in plus 1in\penalty -5\vskip 12pt plus -1in
      {\bf #1}\par\nobreak}





\newdimen\boxmarg 




\def\hline{\noalign{\hrule}}
\def\matrixtable#1{\,\vcenter{\normalbaselines\mathsurround=0pt
   \tabskip=2.5pt\offinterlineskip
   \halign{\hfil$##$\hfil &&\hfil$##$\hfil\crcr
      \mathstrut\crcr\noalign{\kern-12pt}
      #1\crcr\mathstrut\crcr\noalign{\kern-12pt}}}\,}

\def\bivec#1{\vbox{\ialign{##\crcr $\leftrightarrow$\crcr\noalign{
   \kern-1pt \nointerlineskip}$\hfil\displaystyle{#1}\hfil$\crcr}}}


\def\ut#1{\mathop{\vtop{\ialign{##\crcr
     $\hfil\displaystyle{#1}\hfil$\crcr\noalign
     {\kern1pt\nointerlineskip}\hbox{$\hfil\sim\hfil$}\crcr
     \noalign{\kern1pt}}}}}

\def\mincir{\ \raise -2.truept\hbox{\rlap{\hbox{$\sim$}}\raise5.truept	
\hbox{$<$}\ }}								
\def\magcir{\ \raise -2.truept\hbox{\rlap{\hbox{$\sim$}}\raise5.truept	
\hbox{$>$}\ }}								%
\def\undersymbol#1#2{\mathop{\vtop{\ialign{##\crcr
     $\hfil\displaystyle{#2}\hfil$\crcr\noalign
     {\kern1pt\nointerlineskip}\hbox{$\hfil#1\hfil$}\crcr
     \noalign{\kern1pt}}}}}







\def\st{\scriptstyle}


\def\pmb#1{\setbox0=\hbox{$#1$}%
  \kern-.025em\copy0\kern-\wd0
  \kern.05em\copy0\kern-\wd0
  \kern-.025em\raise.0433em\box0}
\def\pmbs#1{\setbox0=\hbox{$\st #1$}%
  \kern-.0175em\copy0\kern-\wd0
  \kern.035em\copy0\kern-\wd0
  \kern-.0175em\raise.0303em\box0}

\def\bfs#1{\hbox to .0035in{$\st#1$\hss}\hbox to .0035in{$\st#1$\hss}\st#1}





\def\Xscr{{\cal X}}


\def\xc{\hbox{\rlap{\hskip 1.5pt\raise .75pt\hbox{--}}$\Xscr$}}


\def\du{\rlap{\raise 6.75pt\hbox{\hskip 3pt$\cdot$}}}
\def\dl{\rlap{\raise 4.5pt\hbox{\hskip 2pt$\cdot$}}}
\def\odu{\rlap{\raise 4pt\hbox{\hskip 3pt$^{\circ}$}}}
\def\odl{\rlap{\raise 1.75pt\hbox{\hskip 2pt$^{\circ}$}}}






\font\rnove=cmr9
\def\mincir{\ \raise -2.truept\hbox{\rlap{\hbox{$\sim$}}\raise5.truept	
\hbox{$<$}\ }}								%
\def\magcir{\ \raise -2.truept\hbox{\rlap{\hbox{$\sim$}}\raise5.truept	%
\hbox{$>$}\ }}								%
\doublespace
\hsize 15truecm
\vsize 22truecm

\blankline 
\blankline 
\blankline 
\blankline 
\blankline

\centerline{{\bf THE COSMOLOGICAL MASS DISTRIBUTION}} 
\centerline{{\bf FROM CAYLEY TREES WITH DISORDER}} 
\blankline
\blankline

\centerline{A. CAVALIERE}
\smallskip
\centerline{Astrofisica, Dipartimento di Fisica II Universit\`a di Roma}
\centerline{via della Ricerca Scientifica 1, I-00133 Roma (Italy)}
\smallskip
\centerline {and}
\centerline{N. MENCI}
\smallskip
\centerline{SISSA, Via Beirut 2-4, I-34014 Trieste (Italy)}
\centerline{permanent address: Osservatorio Astronomico di Roma}
\centerline{Via dell'Osservatorio, I-00040 Monteporzio, Roma (Italy)}
\vskip 8cm
preprint ROM2F/94/25, Ap. J. in press 
\vfill\eject

\noindent
{\it ABSTRACT.} We present a new approach to the statistics of the 
cosmic density field and to the mass distribution of high-contrast structures, 
based 
on the formalism of Cayley trees. Our approach includes 
in one random process both fluctuations and interactions of the 
density perturbations. 
We connect tree-related quantities, 
like the partition function or  
its generating function, to the mass distribution. 
The Press \& Schechter mass function and the Smoluchowski kinetic equation 
are naturally recovered as two limiting cases corresponding to 
independent Gaussian fluctuations, and to aggregation of high-contrast 
condensations, respectively. 
Numerical realizations of the complete random process on the tree 
yield an 
excess of large-mass objects relative to the Press \& Schechter function. 
When interactions are fully effective, a power-law distribution with
logarithmic slope -2 is generated. 
\blankline
{\it Subject headings:} cosmology: theory -- galaxies: clustering -- 
large scale structure of the universe -- galaxies: formation 
\blankline

\smallskip\noindent
{\it 1. INTRODUCTION}
\smallskip

The mass distribution $N(M)$ of high-contrast structures 
constitutes a central link between observed extragalactic sources
and the physics of the early universe. 

In the canonical scenario such structures form by 
direct hierarchical collapses (DHCs).
In the expanding Universe small overdensities 
above the decreasing background density $\rho$ 
are weakly gravitationally unstable (see Peebles 1993). 
 The contrasts 
$\delta \equiv \Delta\rho/\rho$ grow slowly in their linear regime,  
with $\delta  \propto t^{2/3}$ in a critical universe. 
As the contrasts approach unity the perturbations detach from 
the Hubble expansion, turn around, and in a comparable time non 
linearly collapse  
to end up in high-contrast virialized structures. 
 
As for the ``initial'' conditions, the 
physics of the early universe (see Kolb \& Turner 1990) 
suggests the perturbations started  
in the form of 
a Gaussian random field. Their power spectrum at 
$z < 10^3$ may be piecewise approximated 
by $|\delta_k|^2 
\propto k^n$
with $-3 < n< 1$.  
So nonlinear conditions $ k^3 |\delta_k|^2 \sim 1$ are 
reached sequentially at larger and larger sizes, 
with the rich clusters (Abell 1958) forming now. 

For the range of masses $M \magcir 10^{12} M_{\odot}$ 
the collapses are little affected by dissipation, and 
the DHC scenario gives rise yo 
a theory of elaborate elegance 
(see Peebles 1965; 
Gunn \& Gott 1972;  
Press \& Schechter 1974, hereafter PS; Rees \& Ostriker 1977; 
Bond et al. 1991).
The {\it linear} density field $\delta$ is smoothed or averaged 
at each point with a filter function of effective size $R$ corresponding 
to a mass $M\propto \rho R^3$. 
On each scale $M$, the variance of $\delta(M)$ yields the dispersion 
$ \sigma \propto M^{-a}$ ($a \equiv 1/2 +n/6$) of the 
``initial'' Gaussian distribution $p(\delta, \sigma)$.  
The {\it nonlinear} collapses are modeled 
after the pattern provided by  the  top-hat smoothing  filter; namely,  
isolated spherical and
homogeneous overdensities  
which virialize, with a definite characteristic  mass
 $M_c(t) \propto t^{2/3a}$ in a critical universe,  
by the time when the actual contrasts reach values $\sim 2~10^2$.    
As this time would correspond to  a nominal 
contrast $\delta_c \approx 1.7$ 
in the extrapolated linear behavior, such 
threshold 
is taken to separate the linear regime from bona fide condensations. 

The condensed mass fraction is provided by the fraction of spheres 
where the linear $\delta (M)$ crosses the threshold $\delta_c$. 
The mass assigned to each collapsed object 
is twice that  of the largest sphere wherein such a condition applies, 
while the smaller ones are disregarded. 
This elaborate selection rule can be written simply as $N(M)\,M\, dM = 
-\,2\rho\,d\int_{\delta_c}^{\infty} d\delta p(\delta, \sigma)$, as
originally proposed by PS, and in terms 
of $m\equiv M/M_c(t)$ and $\sigma \propto m^{-a}$ yields 
$$N(M,t) = \rho\; {2a \delta_c\over \sqrt{2 \pi} M_c^{2}(t)\sigma} ~m^{-2} 
e^{-\delta_c^2/2\; \sigma^2}~. \eqno(1.1)$$ 

The single parameter $\delta_c$ should comprise 
the complex nonlinear dynamics of the collapses. 
Once its value is set on the basis of, e.g., the top-hat model, 
the functional form of eq. 1.1 at any given $t$ self-similarly depends 
on the initial spectrum only; 
specifically, on 
the spectral index $n$ and on the amplitude taken by $\sigma$ at the scale 
$8h^{-1}$ Mpc singled out by unit variance in galaxy counts. 
Self-similarity is stressed in the analysis 
of Bond et al. 1991, who 
examine the linear field at {\it one} 
epoch (say,  the present $t_o$)   
on the 
resolution $S  \propto M^{-2a} \propto \sigma^2$, 
and compare the resulting  
$ \delta(S)$ 
with a  lowering threshold $1.7~ (t/t_o)^{-2/3}$.

The same authors clarify in terms of excursion sets 
the statistics underlying eq. 1.1.
They show that $\delta (S)$, when extracted from the Gaussian field using 
a sharp k-space filter, 
executes a simple (Markovian) random walk governed by a diffusion equation. 
The diffusive flux density 
per unit  resolution 
of such trajectories, 
having their first up-crossing through the 
threshold $\delta_c$ 
within $dM$ at $M$, 
yields the PS expression complete with shape and amplitude,  
in agreement with the selection rule. 

For all its background, the result 1.1 
has a number of drawbacks. 
The above prescriptions avoid  
overcounting substructures and imply a uniform 
timescale $3t/2$ for all collapses. By the same token, however, they 
are likely to underplay substructures and to  
overestimate the normalization. 
In fact, computations which assign definite 
 sizes and timescales  to the collapses (Cavaliere, Colafrancesco 
\& Scaramella 1991) yield 
longer permanence of 
substructures and a lower normalization 
for $N(M,t)$. 
Equivalently, filters localized in the ordinary rather than in the 
k-space (Bond et al. 1991) change the mass assignment to the 
collapsing density peaks and yield different statistics 
and generally straighter shapes. 
Observations, especially in X-rays, image 
abundant substructures 
within clusters, ranging up to truly binary configurations 
(see Jones \& Forman 1992; White, Briel, \& Henry 1993).

In addition, a considerable body of evidence indicates that 
the direct collapse scenario is 
{\it incomplete} at the high-mass end. 
For example, 
at galactic scales the cD's outnumber the Schechter luminosity distribution (Schechter 1976; Bhavsar 1989); 
the building up of their bodies 
is best understood in terms of 
aggregations of normal members in groups 
as described by the Smoluchowski kinetic equation 
(Cavaliere, Colafrancesco, \& Menci 1992, hereafter CCM). 
Galaxy interactions and merging involving 
one bright partner 
are also likely to stimulate the emissions from active galactic nuclei  
at $z \mincir 2.5$, as indicated by 
statistics, by morphologies of the host galaxies, and 
by the richness of their environments (see  Heckman 1993; Bahcall \& 
Chokshi 1991); numerical experiments and theoretical studies 
also concur to this view (Shlosman 1990; Barnes \& Hernquist 1991). 
At larger scales, the imaging X-ray observations referred to above
provide many snapshots of  groups and clusters at various stages 
of essentially binary aggregations. 

In fact, high-resolution N-body experiments
(e.g., Brainerd \& Villumsen 1992; Katz, Quinn, \& Gelb 1993;  
Jain \& Bertschinger 1994) 
show structure formation to be a far more complex process 
than envisaged by the simple DHC scenario. 
It includes, in addition to direct collapses, frequent 
encounters and aggregations within 
clusters and within larger scale, precursor 
structures with the dimensionality of sheets and filaments. 
The resulting $N(M,t)$ shows, relative to eq. 1.1, 
 a slower evolution 
and a different shape due to an excess at large $M$. 

Here we propose a novel approach 
to a satisfactory $N(M,t)$. 
We discuss in \S 2 and \S 3
how the mass distribution is generated by a {\it complete} statistics 
in the resolution-contrast plane. This comprises as limiting cases both 
the direct collapses from independent 
Gaussian fluctuations (\S 4), and
pure dynamical aggregations of high-contrast condensations
(\S 5). 
The 
competition and mixing between these two components is  computed in \S 6, 
with a net outcome depending  on the mass range and 
on ambient conditions. 
This balance is of keen 
interest because interactions are likely to contribute, as noted  above, 
key common features of apparently diverse astrophysical phenomena 
in the nearby and in the distant Universe. 

\blankline
{\it 2. DISORDER AND BRANCHING}
\smallskip

The language of Cayley trees with disorder (see Derrida \& Spohn 1988)
conveniently describes at the ``microscopic'' 
level of density contrasts $\delta$ how new collapses 
from the linear perturbations 
compete or combine with aggregations of the existing condensations. 

The tree is a computational structure (visualized by fig. 1) 
where at each step $\mu$ random weights $w_i$ are  
 extracted in a cascade following a sequence of links, 
which may  randomly branch into two. 
As the generation number $\mu$ increases, 
the progressive product of such random weights $w_i$ will be related 
to the probability of finding a fluctuation of the density field. 
The tree coordinate $\mu$ will be 
related to the mass scale, and the probability will be related to the number 
of condensations per unit mass. 
The end result 
at the ``macroscopic'' level 
will be a mass distribution $N(M,t)$. 

The tree includes in one random process the following two 
components of the $\delta$ field: 
(1) disorder - with increasing resolution $S \propto \sigma^2
\propto M^{-2a}$ the independent values taken at a given time by 
$\delta(M)$ execute, as recalled in \S 1, a pure {\it random walk}; and 
(2) branching - $\delta$ may also jump by 
stochastic ``{\it branching}'', actually coalescing two paths 
into one. 

The statistics of the combined process is conveniently derived 
in terms of the partition function computed at each 
generation  $\mu$ along the tree in the direction 
of coalescence from an initial  $\mu_o$: 
$Z_{\mu}= \sum_{[paths]}\Pi_{i=\mu_o}^{\mu}\, w_i$, 
where the product refers to the $i$-th preceding 
serial links of the tree, and the sum includes all paths coalesced 
at $\mu$. 
The distribution function of $Z$ will be $P_{\mu} (Z)$, with moments 
$$\langle Z^k \rangle_{\mu} \equiv 
\int dZ\; P_{\mu}(Z)\; Z^k ~.\eqno(2.1)$$ 
It proves technically convenient to set $w_i = 1+v_i$, 
to exhibit the unbiased average $\overline{w}_i=1$  when $v$ fluctuates 
around 0. In this representation 
the partition function reads
\footnote{$^{(1)}$}{\rnove
For $v\rightarrow 0$, the limit we shall consider, 
this is indistinguishable from the 
other representation $Z_{\mu}= \sum_{[paths]} e^{-\sum_{i=\mu_o}^{\mu} v_i}$. 
The latter is heuristically attractive because one may directly identify  
$\sum_i v_i$ with  $\sum_i\delta_i=\delta$ and use the relationship 
$\delta = E/E_b$ 
with the energy $E$ of linear perturbations in a critical universe  
normalized to the background potential energy (Peebles 1980). 
The resulting 
$Z_{\mu}= \sum_{[paths]}  e^{-\sum_{i=\mu_o}^{\mu} E_i/E_b}$ 
has the form of a standard partition function, 
and the counting 
expressed by the definition (2.2) implies a sum rule for the linear energies. 
Nonlinear superpositions of comparable energy fluctuations are explicitly 
accounted for by $\sum_{[paths]}$ in the definition (2.2) and
corresponding to the relation (2.3b).  }
$$Z_{\mu}= \sum_{[paths]}\Pi_{i=\mu_o}^{\mu}\, (1+v_i)~. \eqno (2.2)$$

For the specific tree in fig. 1 the definition of $Z$ 
implies the following recursion relations to hold at each elementary step $d\mu$:

$$Z_{\mu + d\mu}=\left\{
\matrix{ 
(1+v)~Z_{\mu}\hfill & ~~~~with ~ probability ~ 1-\eta d\mu, \cr 
(1+v)[Z_{1\mu}\,u_1 + Z_{2\mu}\,u_2] & 
~~~~with ~ probability ~ \eta d\mu . \hfill \cr 
}
\right. ~ \eqno (2.3)$$
The first line describes 
pure disorder, and the second includes branching; 
$v$ and $u$ are stochastic variables. 
Gaussian initial conditions for the perturbation field imply 
for $v$ a distribution which for small $v$ goes into 
a Gaussian with variance $D\,d\mu$ proportional to the step 
length:  
$$ g(v) = e ^{- v^2/2 Dd\mu}/(2\pi D d\mu)^{1/2}~.\eqno(2.4)$$
Another stochastic function $r(u)$, generally far from symmetric,   
governs the distribution of the weights $u$ 
translating, as we shall show, the interaction probabilities. 

A compact way to embody 
all moments is in terms of the generating function
$$G_{\mu}(x) = \langle e^{-(1+x)Z_{\mu}} \rangle\,. \eqno(2.5)$$ 
Successive derivatives of $G$ at $x=0$ are related to moments of 
$Z$ of increasing order, as shown by the formal expansion 
$$G_{\mu}(x) = \sum_k (-1)^{k}\,(1+x)^{k} \langle Z^k \rangle_{\mu}/k!~~.
\eqno(2.6)$$

Correspondingly, the evolution of the moments can be embodied in a master 
differential equation equivalent to the 
recursion relations eq. 2.3. 
It follows from eq. 2.5 that in an interval $d\mu$ 
the two components simply add with the 
weights provided by their probabilities as given eqs. 2.3, 
and their superposition gives 
$$ G_{\mu+d\mu}(x) = (1-\eta d\mu)\; \overline{G}_{\mu}
+\eta\,d\mu \tilde G_{\mu}^2 ~~~~~ with \eqno (2.7)$$ 
$$\overline{G}_{\mu} \equiv \int dv\, g(v)\, G_{\mu}(x+v+vx), ~~~
\tilde G_{\mu} \equiv \int du \; r(u)\; G_{\mu}(x\,u)~.$$  
The continuous limit $d\mu \rightarrow 0$ 
yields 
$$\partial_{\mu}  G_{\mu}  =  D~\partial_{xx} G_{\mu}/2 +  
\eta(\tilde G_{\mu} ^2 -G_{\mu} )~.\eqno (2.8)$$
For, in this limit 
the lhs yields 
$G_{\mu+d\mu} \rightarrow G_{\mu} + \partial_{\mu}G_{\mu} d\mu$, 
and on rhs the variance of $v$ shrinks proportionally to $d\mu$, so that   
$\overline G_{\mu} \rightarrow   G_{\mu} + 
[D + O(x)]\,d\mu \, \partial_{xx}G_{\mu}/2$. 
When the two sides are set equal, finite terms cancel out,  
and to the first order in $d\mu$ the above equation obtains 
near $x=0$, which will be the relevant point. 

It is easily perceived, and is discussed in detail in \S 3, 
that in the limit of no branching (i.e., $\eta \rightarrow 0$) 
eq. 2.8 reduces to 
a pure diffusion equation for $G_{\mu}(x)$, similar 
to the equation given by Bond et al. 1991 and 
recalled in the Introduction. 
The opposite 
limit of branching with no Gaussian noise yields, as we shall see  in \S 5, 
the Smoluchowski  equation  for the kinetics of $N(M, t)$ under binary 
interactions discussed by CCM.  
We next substantiate 
these two limits by examining the relationship 
of $G_{\mu}(x)$ with $N(M, t)$, and that of the generation 
number $\mu$ with the resolution $S $ or with the physical time $t$. 

We stress that solving eq. 2.8 for $G_{\mu}(x)$ is {\it equivalent} to computing 
the single moments of $Z$ directly from the relations 2.3 and then 
synthetizing 
$G_{\mu}(x)$ from its expansion 2.5. 
The advantage of following this latter route is that the recursion relations 
are very simple,  
and especially suited for numerical work. On the other hand, 
the master eq. 2.8 is 
convenient for making contact with previous work in limiting cases and for 
discussing the balance of the two competing modes. 
The statistical effects of this competition 
are illustrated in fig. 2.

\blankline
{\it 3. FROM THE CAYLEY TREES TO THE MASS FUNCTION}
\smallskip

We first note that in the limit of no branching (i.e., $\eta \rightarrow 0$) 
the remaining terms of eq. 2.8 yield the structure of a diffusion 
equation
$$\partial_\mu G_{\mu}= D\, \partial_{xx} G_{\mu}/2~. \eqno(3.1)$$ 
This, by the gauge (structure-preserving) transformation
$$d\mu = - {1\over 2}\,d~ln\,S\;, ~~~~~~~~ dx^2\propto d\delta^2\;D/ S\;,
\eqno (3.2)$$ 
with the condition $D\,d\mu>0$ (see eq. 2.4), can be made 
identical with the equation (Bond et al. 1991, Lacey \& Cole 1993) 
$$ \partial_S Q= \partial_{\delta \delta} Q/2 ~, \eqno(3.3)$$ 
which governs the evolution of $Q(\delta, S)\; d\delta$, 
the density of trajectories (random walks) of $\delta$ as the resolution 
$S\propto \sigma^2\propto M^{-2a}$ is increased or the scale is 
decreased. 

Because of its central role in what follows, we discuss 
the gauge 3.2 in more detail. 
We first integrate eq. 3.2b to 
$$x \propto (\delta - \delta_c)/\sigma\;, \eqno(3.4a)$$
so that $x=0$ corresponds to $\delta=\delta_c$; 
thus the collapse threshold 
is embedded in the tree formalism as a zero 
point for the tree variable $x$. 

Then we specify (see fig. 3 and its caption) 
the relationship of the remaining 
  tree coordinate $\mu$ and of its initial value $\mu_o$ with 
the {\it physical} variables $M$ and $t$. 
At each time $t$ the resolution scale $S=M^{-2a}$
corresponding to a given generation number $\mu$ is 
reckoned from a minimum 
$S_o \propto [M_c(t)/\epsilon]^{-2a}$ with $\epsilon \ll1$, 
corresponding to 
a maximum mass $M_{max}\equiv M_c(t)/\epsilon\gg M_c(t)$, so 
that eq. 3.2a is integrated in the form 
$$\mu-\mu_o=-{1\over 2}\,ln\,{S\over S_o}=ln\,(\epsilon\,m)^{a}~. \eqno(3.4b)$$
Note that $S_o$ contains 
the $t$-dependence of the initial condition $\mu_o$ for each 
realization of the tree. 
The tree coordinates $\mu_o$ and $\mu -\mu_o$ 
are used as independent ones in what follows. 
The mass scale $M$ is an independent variable in the 
frame of 
the {\it physical} coordinates (the plane $M,t$ in fig. 3), 
but in terms of the {\it tree} coordinates 
becomes a function of $m$ and $t$ through the relation 
$M=m\,M_c(t)$ (see also caption to fig. 3). 

Having so specified the relations between the physical and the tree variables, 
we now elaborate a procedure for computing the mass function from the tree. 
Following the papers by Bond et al. 1991 and Lacey \& Cole 1993, 
the PS selection rule in terms of $Q(\delta,S)$ and $N(m) = N(M) dM/dm$
is written as
$$ N(m)\; m\; dm = - {\rho\over M_c}\,d \int_{-\infty}^{\delta_c}d\delta~ Q 
= - {\rho\over M_c}\,dS \; [\partial_{\delta} Q/2]_{\delta_c}~.\eqno(3.5)$$
The second equality, which expresses the mass fraction as a density 
in resolution of flux across the (absorbing) 
boundary $\delta_c$, is formally provided 
by integration over $\delta$ of eq. 3.3. 
We note that within the above theory 
a Laplace transformation relates $N(M)$ to $Q(\delta, S)$. 
In fact, the integral form 
of the above relation 
may be recovered 
by applying the operator 
$[\partial_{\delta}]_{\delta_c}$ 
to both sides of the integral relation 

$$\int dm\; N(m) e^{-m(\delta-\delta_c)} = {\rho\over M_c}\,\int dS\; Q/2~.
\eqno(3.6)$$
In other words, the selection rule and the diffusion equation imply that 
$ \int dS \;  Q/2$ is the Laplace transform of $N(m)\,M_c/\rho$.
\footnote{$^{(2)}$}{\rnove In the following, unless otherwise specified, 
the integrals are meant to run over the full range of the variables. 
These are as follows: $\delta\in[-\infty,+\infty], \;
m\in[M_{min}, M_c/\epsilon]$, and  $\mu \in [-\infty, \mu_o] $. 
The integrals over $\mu$ ranging in the last interval may be 
rewritten as integrals over  $\mu - \mu_o$ in the corresponding 
range, and are explicitly $t$-independent.}

Based on the close similarity of eqs. 3.1 and 3.3 
when the gauge eqs. 3.2 are considered, we propose 
the following general form of the 
selection rule:  
$${N(m)\over N_T}\, \;m\; dm  = 
-d\,\int dx\,G_{\mu}(x)\,=
- d\mu\; [\partial_x G_{\mu}(x)]_{o}=-d\mu\,
\langle Z_{\mu}\,e^{-Z_{\mu}}\rangle 
. \eqno (3.7)$$
Here $N_T(t) \equiv \int N(m)\,dm\propto \rho/M_c$, and 
the derivative  is computed at 
$x=0$, corresponding to $\delta = \delta_c$
after the gauge transformation 3.4a; 
note from the second term that additive components of $G$ independent of $\mu$ 
will not matter. 
In integral form, the fraction of condensed mass is 
$$ \int dm\;  N(m) ~m = - N_T\,\int d\mu\;[{\partial_x G_\mu}]_{o}~, 
\eqno(3.8)$$
where the last  integral is actually $t$-independent (see footnote $^2$). 
But the above relation also obtains by differentiation $\partial_x$ at $x = 0$ 
of the relation 
$$ \int dm\; N(m)\;e^{-m\,x} = N_T~ \int d\mu\; G_{\mu}(x) 
~,\eqno(3.9)$$
which expresses the generating function, normalized to $N_T$, 
as a Laplace transform of the mass distribution.
The latter then obtains by anti-transforming $G_{\mu}(x)$. 

Equivalently, all successive moments 
of $N(m)$ are obtained by successive differentiation of eq. 3.9. 
For example, the zeroth-order moment is given by 
$\int dm N (m) = N_T~ \int d\mu G_{\mu}(0)$, which fixes the 
normalization $\int d\mu G_{\mu}(0)=1$; the 1st moment is  given 
by eq. 3.8, consistent with the differential form 3.7.

\blankline
{\it 4. DISORDER: THE PS LIMIT}
\smallskip

We now show how the PS mass distribution may be derived 
from eq. 3.7 in the limit of no branching, that is, $\eta\rightarrow 0$. 

The latter equation involves 
$$[\partial_x G_{\mu}]_o = \langle \,Z_{\mu}\; e^{-\,Z_{\mu}} 
\rangle = 
\sum_{k=1} (-1)^{k-1} \langle Z^k \rangle_{\mu}/(k-1)! ~~.\eqno(4.1)$$
The single moments of $Z$ 
will be {\it directly} derived from 
the recursion relation eq. 2.3a (with $D=1$) 
in the form 
$$\langle Z^k \rangle_{\mu+d\mu} = 
\int dv \; {e^{-v^2/2d\mu}\,(1+v)^k \over 
\sqrt{2\pi d\mu}} \langle Z^k \rangle_{\mu} 
~. \eqno (4.2)$$
Expanding the binomial around $v=0$ to the lowest (second) significant order, 
integration of eq. 4.2 yields 
$$\langle Z^k \rangle_{\mu+d\mu} = \Big[1+{k^2-k\over 2}\, d\mu \Big]
\langle Z^k 
\rangle_{\mu}~.\eqno(4.3)$$ 
In the continuum limit $d\mu \rightarrow 0$ this yields  
$$d\; ln \langle Z^k \rangle_{\mu} = {k^2-k \over 2}\, d\mu 
~, \eqno (4.4)$$
which using the gauge 3.2a (i.e., $d\mu=-d\,ln\sigma$) integrates to 
$$\langle Z^k \rangle_{\mu}/\langle Z^k \rangle_o = (\sigma_o/\sigma)^
{k^2-k\over 2}\,.  \eqno(4.5)$$

This relation implies that the first moment of $Z$ is a constant relative 
to $\mu$, which is a natural  consequence of the property $\overline{w}=1$. 
The next two moments yield the leading contributions to the sum 
4.1. In fact, following eq. 4.5 and 
keeping only the $\mu$-dependent terms (as noted 
just after eq. 3.7), we find 
$$\langle \,Z_{\mu}\; e^{-\,Z_{\mu}} \rangle = \langle Z\,\Big[1-Z+
{Z^2\over 2}+...\Big]\rangle = Z_o^2\,{\sigma_o\over \sigma}\Big[1- 
{Z_o\over 2}\,{\sigma_o^2\over \sigma^2}+ O\Big({\sigma_o^5\over \sigma^5}\Big)
\Big]\,.\eqno(4.6)$$
Recalling from \S 3 that  $\sigma_o/\sigma=[M/M_{max}]^a=(\epsilon\,m)^a$, 
it is seen that $\sigma_o/\sigma<1$ holds for $M < M_{max} =M_c/\epsilon$,  
 and the last equality can be resummed to within $O(\epsilon^{5a})$ 
to yield 
$$ [\partial_x G_{\mu}]_o \approx Z_o^2\,{\sigma_o\over \sigma}\,
e^{-Z_o\,\sigma_o^2/2\,\sigma^2}
=Z_o^2\,(\epsilon\,m)^a\,e^{-Z_o(\epsilon\,m)^{2a}}~.\eqno(4.7)$$

The tree by itself, as any statistics,   
 does not specify the dynamics of 
gravitational collapses, and provides only scaling behaviors: these, on 
substituting eq. 4.7 in eq. 3.7, are 
$$ N (m) = {\rho\over M_c} \,m^{-2}\; {d\,ln\,\sigma\over d\,ln\,m} 
~[{\partial_x G_{\mu}}]_o \approx 
 {2\,a\,\rho\,Z_o^2\over M_c}\; (\epsilon\,m)^{-2+a}\; 
 e^{- Z_o\,(\epsilon\,m)^{2a}/2} ~, \eqno(4.8)$$
and turn out to be the same as in the PS distribution. As to the constants 
which do carry dynamical information, 
the initial condition $Z_o$ may be set to $\delta_c$, the counterpart 
here of the boundary condition used by Bond et al. 1991. 
In addition, the rescaling $N(\epsilon\,m)=N(m)/\epsilon$ holds, and  
the overall normalization follows from the requirement 
$\int d\mu\,G_{\mu}(0)=1$ derived in \S3. 
Thus the full PS expression 1.1 is recovered. 

\blankline
{\it 5. BRANCHING: THE SMOLUCHOWSKI LIMIT}
\smallskip

We now consider eq. 2.8 in the opposite limit of branching 
with no Gaussian noise. 
Then the 
remaining terms on rhs read 
$$ \partial_{\mu}  G_{\mu} = \eta(\tilde G_{\mu}^2 -G_{\mu}). \eqno (5.1)$$
Equivalently, the change of single moments of $Z$ may be derived {\it directly} 
from the recursion relation 2.3b with 
no noise, to yield 
$$\langle Z^m\rangle_{\mu+d\mu}=
d\mu\,\int\int dZ_1\,dZ_2\,P(Z_1)\,P(Z_2)\,(Z_{1\mu} +Z_{2\mu})^m $$ 
$$+(1-d\mu)\int dZ\,P(Z)\,Z^m_{\mu}~; \eqno(5.2)$$
here the coupling parameter $\eta$ has been absorbed into a rescaling 
of $\mu$ for convenience as will become apparent after eq. 5.11. 
Expanding the binomial, this becomes 
$$\langle Z^m\rangle_{\mu+d\mu}  
=d\mu \int\int dZ_1d\,Z_2\,P(Z_1)\,P(Z_2)\,\sum_{k=1}^m \,
{m!\over k!\,(m-k)!}\, Z^k_{1\mu}\,Z^{m-k}_{2\mu}$$
$$+(1-d\mu)\langle Z^m\rangle_{\mu}.\eqno(5.3)$$
In the continuous limit $d\mu\rightarrow 0$ this yields
$$\partial_{\mu}{ \langle Z^m\rangle_{\mu}\over m! }=
\sum_{k=1}^m{ \langle Z^k\rangle_{\mu}\over k! }\,
{ \langle Z^{m-k}\rangle_{\mu}\over (m-k)! }
-{ \langle Z^m\rangle_{\mu}\over m!}~. \eqno(5.4)$$

This relation 
has a structure similar to the Smoluchowski equation which   
governs the kinetics of binary aggregations of definite condensations 
(CCM): 
$$ {\partial N \over \partial t} = {1 \over 2} \int_0^M dM'~ 
K (M',M-M',t) ~N(M',t) ~ N(M-M',t) $$
$$ - N(M,t) \int_0^{\infty} dM'~ 
K(M,M',t) N(M',t) ~. \eqno(5.5)$$
The kernel $K$ represents the 
rate of aggregation in a 
system of condensations with relative velocity $V$ and 
interaction cross section $\Sigma$, and averages to 
$\tau^{-1} = N_T \overline{\Sigma V}$ in terms of the component 
number density $N_T(t)$.

Indeed, we show next that the above equation may be 
identified with 
eq. 5.4 (or with the equivalent eq. 5.1) when dealing with 
high-contrast 
condensations insensitive to the perturbation field; that is, 
in the limit of no disorder (with $\delta=\sigma$) and of high-contrast
  condensations (i.e., $\delta_c\rightarrow 0$). 
For a constant kernel $K$ we again proceed directly. 

To this end, we first rewrite the discretized form of eq. 5.5, 
after dividing by $N_T/2$, in the form 
$${2\over N_T^2}\,{\partial N_M\over\partial t}+{N_M\over N_T}
=\sum_{M'} {N_{M-M'}\over N_T}\,{N_{M'}\over N_T}
-{N_M\over N_T}~, \eqno(5.6)$$
where the parameter $K$ has been absorbed 
into a rescaling of $t$ 
for convenience which again will become apparent after eq. 5.11 below.   
We then express the $t$-derivative in terms of the tree variable $\mu$. 
The gauge eq. 3.2a yields 
$d\mu = - {1\over 2} d\, ln S=a\,d\,ln\,M$. In the tree frame 
(as we discussed in \S 3) 
$M=m\,M_c(t)\propto m/N_T$ hold, and one obtains 
$$d\mu \propto - {dN_T\over N_T}={1 \over 2} N_T\, dt~,\eqno(5.7)$$ 
with the last equality coming from eq. 5.5 integrated over $M$. 
Then eq. 5.6 can be written as 
$${\partial\over \partial\mu}{N_M\over N_T}=\sum_{M'} {N_{M-M'}\over N_T}
\,{N_{M'}\over N_T} - {N_M\over N_T}~.\eqno(5.8)$$ 


Now the above equation exhibits even more clearly a structure similar to 
eq. 5.4, and is identical to it provided that 
$${N_{M}\over N_T} \propto {\langle Z^m \rangle_{\mu}\over m!}\eqno(5.9)$$
holds. 
The simplest way to 
prove this relationship is to  to write the integral relation 3.9 
with the shorthand $1+x=w$ 
in the form 
$$\int {N(m)\over N_T}\,w^m\,dm=
\int d\mu \langle 
e^{-w Z_{\mu}}\rangle =\sum_k 
\,w^{k}\,\int 
d\mu \,{\langle Z^k\rangle_{\mu}\over k!}
~, \eqno(5.10)$$
with the signs $ (-1)^k$ 
included into $\langle Z^k \rangle$ 
in view of the 
invariance of eq. 5.4 relative to such transformations. 
Then we note that eq. 5.4 for $m =1$ implies $d\,ln \langle Z \rangle_{\mu} \propto d\mu$, 
and hence $\langle Z \rangle_{\mu} \propto N_T$ by eq. 5.7. For $m=2$ 
it implies $\langle Z^2 \rangle \propto N_T$ asymptotically when 
$N_T \ll N_{To}$ holds; similarly for the higher orders. 
With $\langle Z^k\rangle_{\mu}/N_T \, k! \rightarrow $const relative to 
$\mu$, the integration on rhs of eq. 5.10 reduces to 
$\int d\mu N_T = \int d\mu e^{-\mu}$ in view of eq. 5.7, 
and yields a constant factor. 
Thus, writing the integral on the lhs of eq. 5.10 in discrete form, 
we find 
$$\sum_m N_m\,w^m  \propto \sum_k  
w^{k} \langle Z^k \rangle_{\mu}/k! ~, \eqno(5.11)$$ 
where of course the dummy indexes $k$ and $m$ may be identified. 
We now apply to both members the operator $[\partial^m_w]_{w=0}$, 
which in fact is appropriate for 
high-contrast structures 
with $\delta_c\rightarrow 0$ and $\delta\approx\sigma$.
The result is the relation 5.9, thus completing 
the identification of the Smoluchowski equation of the form 5.5 
with the tree recursion relations of the form 5.4. 

The proportionality constant in eq. 5.11 translates into a rescaling 
by a constant factor of the correspondence between 
the independent variable in eq. 5.4 (which in full is $\mu \eta$),  
and that in eq. 5.6 (which in  full is $t/\tau$). 
The statement that $\sum_m\langle Z^m\rangle_{\mu}/N_T\,m!$ 
becomes time-independent holds for the solutions $N(M,t)$ of the 
Smoluchowski equation in their asymptotic, self-similar stage where 
$N(M)/N_T^2$ is time-independent.

The equivalence of 
the Smoluchowski equation 5.5 with a Cayley tree 
can also be proved when the  kernel $K(M, M')$ is mass-dependent. 
Then the equivalence may be recovered by going through the Laplace 
transform  of eq. 5.5, as has been 
carried out explicitly by Peshanski (1992)
for the case of multiplicative kernels $K(M,M')=K(M)\,K(M')$. 
The link between kernel and weight distribution on 
the tree $r(u)$ 
is explicitly given by 
$$K(m)=C\,\int_0^{\infty}\,du\,r(u)\,e^{um}~, \eqno (5.12)$$
which amounts again to a Laplace 
transformation. The constant $C \propto \tau^{1/2}$ contains the normalization 
 of $K$, since $r(u)$ is normalized to $1$.

\blankline
{6. DISORDER AND BRANCHING SUPERPOSED} 
\smallskip

We now compute numerically the mass function 
in conditions where both disorder and branching are effective. 
As has been said, for numerical work it is much easier and faster 
to use {\it directly} the 
recursion relations (eq. 2.3) for the partition function $Z_{\mu}$ 
--  rather than the master equation 2.8. 

This is done  by the following 
procedure. 
At each step $\mu$ of the tree a numerical 
algorithm extracts the weights $v$ from a Gaussian 
distribution, and uses the first recursion relation 2.3 with probability 
$1-\eta\,\Delta\mu$, or the second with probability $\eta\,\Delta\mu$,  
to construct the partition function at the next step. In the 
present computations 
we use a delta-function $r(u)=\delta(u)$, corresponding after eq. 5.12 
to a constant interaction kernel. 
In fig. 2 we 
have shown the resulting 
trajectories of the contrast for increasing $S$, which are related to the 
tree variables $x$ and $\mu$ by the gauge eqs. 3.4. 

For each tree, we generate the entire $Z_{\mu}$ and its moments. 
The procedure is repeated for a large number of trees (up to $10^3$). 
The moments of the partition function are used to compute 
$$[\partial_x G_{\mu}]_o = 
\sum_{k=1} (-1)^{k-1} \langle Z_{\mu}^k \rangle/(k-1)!,\eqno(6.1)$$
 where the sum we 
actually used runs up to $k=10$. Then at any given $t$ we compute the mass distribution 
$N(M,t)$ following eq. 3.7, and check that the total mass is conserved in 
time to within 1\% 
when $10^3$ trees are used. 

The result depends on the branching probability 
$\eta$, which acts like an effective coupling constant for 
binary interactions. For small $\eta \mincir 10^{-2}$ the diffusion part of 
eq. 2.8 
dominates and one recovers the PS mass distribution. 
 For increasing values of $\eta$ the resulting distributions become 
steeper, while the cutoff at large masses tends to straighten up. 
When $\eta\sim 10^{-1}$ the distribution reaches its {\it asymptotic} form, 
a pure power law with logarithmic slope close to $-2$; see fig. 4

There are two reasons why even a small value of $\eta$ is effective. 
First, note that on transforming eq. 2.8 following the gauge eqs. 3.4 
the effective coupling parameter is 
$\eta/S \propto \eta m^{2a}$,  
that is, a coupling more effective at 
larger masses. 
Second, a kernel constant in time, as considered here, maximizes 
the interactions because it implies 
no dependence of the number density on ambient evolution. 
This is physically realistic for interactions within large scale 
structures surviving for longer then $\tau$. 

These results compare interestingly with the results from 
  cosmological  
N-body simulations with a large dynamical range and highly resolved 
data analysis, 
like those performed by Brainerd \& Villumsen 1992 and 
Jain \& Bertschinger 1994.  
These papers agree in finding a slower evolution and an excess at large masses
compared with the PS mass distribution,  
with the former finding consistency with 
a simple power law distribution, approximatively $M^{-2}$ within the 
walls and filaments. 
These features are consistent  with our findings. 

In terms
of overdensities vs. resolution 
as used in fig. 2, the interactions cause not only an obvious density 
decrement of the trajectory distribution toward larger coalesced masses, 
but also a skewness relative to the Gaussian counterpart. 

\blankline
{\it 7. DISCUSSION} 
\smallskip

The existing theories for the shape and the evolution 
of the mass distribution $N(M,t)$ 
fail to capture 
the full complexity of cosmic structures. 
The PS theory, with its recent elaborations, provides with eq. 1.1 
the best quantification for 
the scenario envisaging hierarchical collapses from initial density perturbations.  
Yet the result differs  as to amplitude, 
shape, and evolution from data or from simulations and from 
realistic excursion set computations 
with the same input parameters.
The alternative mode to hierarchically building up structure is 
based on 
binary aggregations between high-contrast condensations. This  
predicts 
non-Gaussian formation of rare large objects, and 
describes more closely the erasure of substructures within a structure by 
resolving timescales different from its dynamical time $t_d$. 
But it requires input, 
at least initially,  
of formed condensations into an environment protected from 
the Hubble expansion (CCM).

We submit that both these modes constitute only 
partial representations of the evolution of the density 
field. They  select either purely Gaussian 
random walks of the linear density contrasts 
as functions of the resolution, 
up to crossing the  threshold of nonlinearity; 
or  trajectories  ``branching'' stochastically 
only at 
large values of the contrast. 
Correspondingly, the PS  function 1.1 and the Smoluchowski 
aggregation  equation 5.5 correspond to {\it restricted}, 
``macroscopic'' 
averages from a {\it complete} 
`` microscopic'' field statistics
which treats aggregations and 
direct collapses as proceeding together at similar contrast levels. 

We propose such a complete statistics  which combines, 
in the form of a Cayley tree 
(or random cascade) as represented in fig. 1, random walk 
and stochastic branching  of 
fluctuations into one partition function governed  
by the single {\it master} equation 2.8. We also propose 
the Laplace transform relationship eq. 3.8 between 
 the tree generating function $G_{\mu}(x)$ and  the mass distribution 
$N(M,t)$. 

We have proved that our proposals indeed yield in the appropriate 
limits both the diffusion equation 
(\S 4) with the associated PS function 
eq. 1.1, and the Smoluchowski equation (\S 5). The former limit 
applies to Gaussian fluctuations 
that {\it independently} reach the nominal threshold 
for collapse and virialization. The latter limit 
is derived with no formal recourse to a threshold, 
and describes binary {\it interactions} of an initially given 
set of already formed condensations. 
Pleasingly, from averaging  over the tree 
in the latter limit one  obtains a mean-field, kinetic {\it equation} 
with different solutions depending on the  
kernel, and evolving away from the initial form; in the former 
limit, one instead obtains a single PS {\it function}.

Both limiting processes are hierarchical and imply ``merging'', i.e., 
inclusion of smaller condensations into larger ones.
But pure DHC in fact envisages only reshuffling 
of (generally) many condensations which belong to lower hierarchical levels 
into a higher level on a {\it larger} mass scale at a 
{\it subsequent} time, strictly following the 
conditions 
set ab initio in the linear fluctuation field.
 The natural quantification of this process 
is provided by the conditional probability that a trajectory 
up-crossing the threshold at the resolution $S_2$ had a previous 
up-crossing at $S_1 > S_2$ (see Bond et al 1991; Bower 1991; Lacey \& Cole 1993).

Pure aggregation, on the other hand, envisages 
pairs of condensations {\it coalescing} into a third,  
at a {\it similar} -- and high --  
contrast level. 
Here the nonlinear interactions are 
stochastically
set by ambient density and relative velocities, 
and may be triggered
at any time when larger scale structures outline volumes with 
 expansion slowed down relative to the general ``field''.

The Cayley tree approach not only proves successful in deriving the 
two, oppositely extreme modes for hierarchically  
building up structures, 
but also provides a number of links between them: 
the common tree algorithm or the master equation 2.8; 
the Laplace transform relationship equation 3.8 
of $G_{\mu}(x)$ with $N(M,t)$, 
which extends the PS selection rule to interacting fluctuations; 
and the tree variable  $\mu$ that 
includes 
both the resolution $S$
(used to derive the PS function) and the 
physical time $t$ (used to derive the Smoluchowski equation), as 
stressed below. 

In fact, we have seen that 
$d\mu = - {1\over 2}\,d\, ln S$
holds, appropriate to 
the former case. On the other hand, 
from the relation 
$M=m\,M_c(t)\propto m/N_T$ one obtains $\partial_t\mu \propto 
-d\,ln N_T/dt$, which is the eq. 5.7 
used in deriving the Smoluchowski equation.
The constant factors (and specifically 
the spectral index $n$, 
the last remnant of the initial perturbation spectrum) 
are actually irrelevant in the 
subsequent identification of eq. 5.5 with the tree recursion 
relation in the absence of Gaussian noise, as expected. 

Above all, the tree formalism describes with equal ease 
the mixing and competition 
of the two pure modes; that is, it describes 
condensations interacting and 
aggregating while still growing or collapsing. The process may be computed 
either from the master equation, or directly in terms of the tree partition 
function, which is very convenient and fast on computers. 
These computations yield, when branching is fully effective,  
$N(M,t)$ in the form of a steep, 
slowly lowering power law  with logarithmic slope $\approx -2$.
The result is due to 
the continuous input from the Gaussian noise over many scales, 
combined with the increasing effect of the branching mode at larger masses. 

In fact, even a value of $\eta<1$ is effective because  
the relevant coupling parameter 
$\eta/S \propto \eta\,m^{2a}$ increases at 
higher masses. 
Actually, the interactions are maximized 
by a kernel constant in time as considered here, corresponding to 
densities and velocities unaffected by 
expansion. This is physically realistic for 
interactions taking place within large-scale 
structures which survive for times 
longer that $\tau$. 
In these conditions 
high-mass condensations form faster than the {\it Gaussian} rate 
within large scale structures, not unlike the formation action 
seen in large N-body
simulations (Brainerd \& Villumsen 1992; Babul \& Katz 1992). Observations 
show that groups are concentrated in filaments 
and sheets outlined by redshift surveys of galaxies (Ramella et al. 1990) 
and that the mass distribution over scales from $10^{13}~ M_{\odot}$ to 
several $10^{15}\,M_{\odot}$ 
is consistent with a steep power law (Giuricin et al. 1993). 

We  shall discuss elsewhere the relationship of our approach 
with the dynamical descriptions of matter field under gravity -- e.g., the 
adhesion model (see Shandarin \& Zel'dovich 1989) or the 
frozen-flow approximation (Matarrese et al. 1992) -- which include shear 
effects in addition to interactions of comparable fluctuations. 
Conditions of aggregating interactions protected from the Hubble expansion 
within larger structures  
are likely to underlie apparently diverse 
astrophysical phenomena, from formation of one large  
cD-like galaxy by strong interactions of members in groups, 
to the very formation and growth 
of groups and clusters within large scale filaments and sheets.
Toward such a complex emergence of cosmic structure   
Cayley trees provide a unifying approach. 
\smallskip 
\noindent
Acknowledgments. It is a pleasure to thank Margaret Geller for stimulating 
exchanges,  
and Felix Ritort for helpful discussions. We acknowledge financial support 
by ASI, MURST and the EC HCM Programme.

\blankline
\blankline

\vfill\eject

\noindent
{\it REFERENCES} 
\smallskip
\ref Abell, G.O. 1958, ApJS., 3, 211 
\ref Babul, A., \& Katz, N. 1992, ApJ, 406, L51
\ref Bahcall, N.A., \& Chokshi, A. 1991, ApJ, 380, L9
\ref Barnes, J.E., \& Hernquist, L.E. 1991, ApJ 370, L65
\ref Bhavsar, S.P. 1989, ApJ, 338, 718
\ref Bond, J.R., Cole, S., Efstathiou, G., \& Kaiser, N. 1991, ApJ, 379, 440
\ref Bower, R.J. 1991, MNRAS, 248, 332
\ref Brainerd, T.G., \& Villumsen, J.V. 1992, ApJ, 394, 409
\ref Cavaliere, A., Colafrancesco, S., \& Menci, N. 1992, ApJ, 392, 41 (CCM) 
\ref Cavaliere, A., Colafrancesco, S., \& Scaramella, R. 1991, ApJ, 380, 15
\ref Derrida, B., \& Spohn, H. 1988, J. Stat. Phys., 51, 817
\ref Giuricin, G., Mardirossian, F., Mezzetti, M., Persic, M., \& Salucci, P.
1993, SISSA preprint
\ref Gunn, J.E., \& Gott, J.R. 1972, ApJ, 176, 1
\ref Heckman, T.M. 1993, STScI preprint  
\ref Jain, B. \&  Bertschinger, E. 1994, MIT preprint
\ref Jones, J.C. \& Forman, W. 1992, in Proc. NATO-ASI 
{\it Clusters and Superclusters of Galaxies}, 
ed. A.C. Fabian (Dordrecht: Kluwer), p. 49
\ref Katz, N., Quinn, T. \& Gelb, J.M. 1993, MNRAS, 265, 689
\ref Kolb, E.W., \& Turner, M.S. 1990, {\it The Early Universe} 
(Redwood City, CA: Addison-Wesley)
\ref Lacey, C., \& Cole, S. 1993, MNRAS, 262, 627
\ref Matarrese, S., Lucchin, F., Moscardini, L., \& Saez, D. 1992, 
MNRAS, 259, 437
\ref Peebles, P.J.E. 1965, ApJ, 142, 1317
\ref ------. 1980, {\it The Large Scale Structure of the 
Universe} (Princeton: Princeton Univ. Press)
\ref ------. 1993, {\it Principles of Physical Cosmology} (Princeton: 
Princeton Univ. Press)
\ref Peshanski, R. 1992, CERN preprint TH.6044/91
\ref Press, W.H., \& Schechter, P. 1974, ApJ, 187, 425
\ref Ramella, M., Geller, M.J. \& Huchra, J.P. 1990, ApJ, 353, 51
\ref Rees, M.J, Ostriker, J.P. 1977, MNRAS, 196, 381
\ref Shandarin, S.F \&  Zel'dovich, Ya.B. 1989, Rev. Mod. Phys., 61, 
185 
\ref Schechter, P.L. 1976, ApJ, 203, 297
\ref Shlosman, I. 1990, in IAU Colloq. 124, {\it Paired and Interacting 
Galaxies}, ed. J. Sulentic \& W. Keel (Dordrecht: Kluwer), p. 689 
\ref von Smoluchowski, M. 1916, Phys. Z., 17, 557
\ref White, S.D.M., Briel, U.G., \& Henry, J.P. 1993, A\&A, 259, L31 

\vfill\eject

\newline
{\it FIGURE CAPTIONS}{}

\blankline
\noindent
Fig. 1. A schematic representation of a Cayley tree. The random process is 
represented as a cascade proceeding from the bottom up. 
At each step  (labeled by the generation number $\mu+d\mu$) 
a random weight $w =1+v$ is extracted according to a Gaussian
distribution $g(v)$. In addition, the path may either 
result 
from coalescing (with probability $\eta\,d\mu$) 
of two branches that were still distinct at $\mu$, 
or proceed along a single branch. 
The partition function at a point $\mu+d\mu$ is constructed by summing the 
products of the 
weights over all paths leading to it. 
The heavy line marks a specific path on the tree with 
weights $w_1$, $w_2$, ... $w_7$.
\blankline
\noindent
Fig. 2. Contrast vs. resolution (in arbitrary units) of a number of trajectories  
computed from 
numerical realizations of the tree (details are given in \S 6). 
{\it Top panel}: Pure Gaussian noise resulting in 
random walks with dispersion increasing with increasing resolution 
(or decreasing 
mass). {\it Bottom panel}: Effects of including random branching 
(with $\eta=0.1$) 
in the statistics of the trajectories; the branching points are marked by a 
dot. The study of the apparent differences 
between the two panels constitutes the thrust of this paper. 
\blankline
\noindent
\noindent
Fig. 3. Relations between  the physical variables 
$M$ and $t$ and the tree coordinates $\mu$ and $\mu_o$ . At a given $t$ 
({\it top panel}), the initial value $S_o$ for the resolution 
corresponds to a maximum mass scale $M_{max}=M_c(t)/\epsilon\gg M_c(t)$.
It also corresponds to a value of 
the tree generation number $\mu_o \propto ln M_{max}$ following eq. 3.2. 
So $\mu-\mu_o\propto ln\,m $ holds, with $m \propto M/M_c(t)$.
For a given mass $M$ ({\it bottom panel}), 
different values of $t$   
correspond to various starting points $\mu_o$,  and to 
different values of $\mu$ 
visualized by the intercepts of a horizontal line for 
increasing $t$. \blankline
\noindent
Fig. 4. Mass function $N(M)$ resulting from numerical realizations of trees
with both branching and disorder. The partition function and the 
related moments 
are computed from the eq. 2.3. 
{\it Dotted curve}: $\eta =0$; {\it dashed curve}: $\eta = 0.05$; 
{\it solid curve}: $\eta = 0.1$. The first 
is identical with the Press \& Schechter function, eq. 1.1.
\blankline
\noindent
\bye